\documentclass[conference,letterpaper]{IEEEtran}
\IEEEoverridecommandlockouts
\usepackage{cite}
\usepackage{amsmath,amssymb,amsfonts}
\usepackage{graphicx}
\usepackage{textcomp}
\usepackage{xcolor}
\usepackage{subcaption}
\usepackage{soul}
\usepackage[normalem]{ulem}
\usepackage{url}
\sethlcolor{yellow}
\usepackage{fancyhdr}
\usepackage[ruled,vlined,linesnumbered]{algorithm2e}
  \SetCommentSty{textit}
  \SetKwComment{Comment}{$\triangleright$\, }{}
  \SetKwFunction{FDetect}{DetectAnomaly}
  \SetKwFunction{FEval}{EvaluateCell}
  \SetKwFunction{FChildren}{GetChildren}
  \newcommand{\Var}[1]{\textit{#1}}
\usepackage[draft]{hyperref}

\newif\ifcomments
\commentstrue

\definecolor{darkgreen}{RGB}{0,95,0} 
\definecolor{ultramarine}{rgb}{0.07, 0.04, 0.76}
\definecolor{brickred}{rgb}{0.8, 0.25, 0.33}

\begin{document}

\title{Lightweight Multi-scale Hierarchical Anomaly Detection and Localization for Geospatial Big Data Applications at the Edge}

\author{Thomas~Benton~Townsend\textsuperscript{1}, Joshua~Bean\textsuperscript{1}, Benjamin~K~Tkach\textsuperscript{2}, Narcisa~Gabriela~Pricope\textsuperscript{3}, \\ and~Dimitrios~Michael~Manias\textsuperscript{1} \\ 
The Department of Computer Science and Engineering\textsuperscript{1},  
The Department of Political Science and Public \\ Administration\textsuperscript{2}, and 
The Department of Geosciences\textsuperscript{3}\\ 
Mississippi State University\\
\{tbt135, jab1896, b.tkach\}@msstate.edu, npricope@research.msstate.edu, dmanias@cse.msstate.edu}

\maketitle

\begin{abstract}
As an increasing number of critical applications, including environmental, emergency, meteorological, and agricultural, rely on real-time anomaly detection in geospatial data streams, challenges related to the storage, processing, and communication of this data arise. Traditionally, large volumes of data have been sent to centralized processing locations for insight extraction. Given the big data context of these applications, this approach becomes increasingly infeasible as data volume and velocity continue to increase. This paper proposes a lightweight edge-oriented approach for anomaly detection and localization for geospatial data streams. By leveraging the H3 discrete global grid system and a multi-scale drill-down logic, the proposed approach significantly reduces computational overhead, achieving a 99.7\% reduction in evaluations compared to traditional flat-scan methods.  Furthermore, by filtering out noise-induced flickering anomalies at lower resolutions, spatially-persistent anomalous signals can be efficiently identified. The results demonstrate that the proposed framework effectively distills massive geospatial data into actionable insights.

\end{abstract}


\section{Introduction}
Geospatial data can be defined as information that connects a measurable attribute to a physical geographical location. In today's world, geospatial data collection can occur using entities such as satellites, planes, drones, terrestrial sensors, and even crowdsourcing. These collection methods lead to a variety of measurable attributes, for example, environmental condition indices, atmospheric quantities, and even societal indicators. Various use cases and applications can be derived from this data, including long-term modeling and condition forecasting, agricultural optimization, and even security and threat detection. To this end, the collection, processing, insight extraction, and communication related to data sources are critical to tapping into their true value and potential \cite{lee2015geospatial}.

Given the multi-modality of geospatial data, its processing varies by source. In the context of satellite-collected data, after collection, raw data is traditionally sent to terrestrial stations for processing, due to the limited on-board computing resources. Other collection methods, such as drones, generally store raw data on board during the flight and process it after the flight concludes. Low-power sensors generally lack the communication capabilities to transmit data over cellular networks and often rely on intermediate gateways, acting as middleboxes, for some level of processing before sending to a centralized location. As the volume and velocity of geospatial data collection scale exponentially, significant challenges arise related to the communication efficiency and security of traditional geospatial processing methods.

The first main challenge relates to the communication resources required to transmit large volumes of raw data \cite{dritsas2025remote}. When considering data collection from low-orbit satellites, a few critical communication challenges emerge. Firstly, these satellites face communication availability challenges and can only transmit data when entering the range of a terrestrial station (transmission window). This suggests that if a transmission window is unavailable and the on-board data storage has been exhausted, further collection (without the deletion of existing data) is not possible. Furthermore, even when a transmission window is available, the amount of data collected (especially when increasing spatial resolution) takes a significant amount of time to transmit, despite using high-frequency bands. Additionally, as more and more data-collection satellites are deployed, the availability of communication resources per satellite decreases, since the frequency spectrum is finite. 

The second main challenge relates to the security of transmitting raw data for critical applications \cite{wu2024geospatial}. The data collected from these satellites is used in applications such as environmental condition monitoring, weather forecasting, disaster response, and agricultural intelligence. A malicious actor undermining the integrity of transmitted data can have devastating consequences for critical decision-making that directly impacts public safety. Securing data transmissions over these communication channels also requires significant on-board processing and computation, and is subject to environmental conditions (such as radiation) that can degrade transmission quality and actively interfere with genuine data decryption efforts. Considering these two identified challenges, circumventing raw data transmission to a centralized location and moving towards a distributed processing-at-the-edge approach is essential.

Many of the critical applications identified above rely on the detection of data anomalies to identify areas of concern requiring further investigation, continued monitoring, or immediate action. There are two main components to anomaly detection: statistical identification and domain knowledge contextualization. These two components work in synergy; the statistical identification alerts to the presence of a potential anomaly, whereas the domain knowledge provides meaningful context to not only validate the presence of the anomaly but also explain its physical implications. 


Anomaly detection methods in geospatial data sources face unique challenges that are inherent to highly dynamic systems. This is especially true for high spatial resolution data that can provide a very detailed view of a system attribute. When determining anomalies at high resolution, there is a risk pertaining to the truthfulness of the anomaly; while attribute measurements may be precise, their interpretation can be flawed as they may be subject to sampling bias. Another challenge pertains to distinguishing between macro anomalies and flickering anomalies. Macro-anomalies are spatially rooted significant events that are observable at various resolutions and demonstrate system-level concerns. Flickering anomalies on the other hand are visible at higher resolutions and get averaged out at lower resolutions. The manifestation horizon of the anomaly is another key consideration. Some anomalies are sudden, with an immediate impact, whereas others gradually emerge over time. Finally, the duration of the anomaly is another key characteristic that must be considered. An anomaly can be described as being instantaneous (short-lived), sustained (persistent), or periodic (seasonal/cyclical). The ability to detect and characterize anomalies in large-scale dynamic systems is critical for the success of the critical applications that rely on them.

To this end, the work presented in this paper presents a hierarchical approach to macro anomaly detection and localization at the edge for geospatial data applications. The presented algorithm aggregates observed quantities into low-resolution spatial groupings and uses thresholding-based methods, built using historical observations, to determine if an anomaly is present. If detected, a drill-down approach is taken to begin localizing the source of the anomaly at the higher resolution spatial groupings. By reducing the unnecessary computational burden by only applying drill-down logic to spatially and statistically significant anomalies, this work presents a lightweight edge-feasible approach to anomaly detection. This work presents a first key step in developing a communication-efficient and secure framework that is insight-driven and does not require the transfer of massive volumes of raw data from a collection source to a centralized destination.

The remainder of this work is structured as follows: Section II establishes the state of the art and introduces related work. Section III presents the system model and establishes the domain as a big data application. Section IV outlines the methodology and experiment design. Section V presents and analyzes the results. Finally, Section VI concludes the paper and presents avenues for future work in the field.

\section{Related Work}

The field of anomaly detection has seen great progress in recent years, with applications ranging from real-time condition monitoring of critical systems \cite{8859386, 10901699} to the detection of network threats and intrusions \cite{falconc, violos2025olida}. Anomaly Detection in geospatial data streams is becoming increasingly important for enabling critical agricultural health and monitoring applications. As such, various methods have been proposed to achieve effective anomaly detection in such data sources. Reshetova \textit{et al.} \cite{reshetova2023semand} propose a self-supervised anomaly detection framework in multimodal geospatial datasets. The authors propose a domain-specific data augmentation strategy, which introduces synthetic anomalies to enhance training. Their detection framework incorporates a two-stage anomaly scoring leveraging a neural network backbone and an out-of-distribution detection method. Yogarajan \textit{et al.} \cite{yogarajan2025hybrid} propose a hybrid approach for geospatial anomaly detection in sensor measurements. The authors leverage isolation forests and density-based clustering to identify various anomaly types. The authors also use spatial correlation indices to determine the probable source of a detected anomaly. Budgaga \textit{et al.} \cite{budgaga2017framework} propose a framework for scalable real-time anomaly detection over voluminous geospatial data streams. In this work, the authors leverage geohashing to partition incoming data streams. The authors evaluate various model types, including density, distance, Bayesian, and ensemble-based. The authors also include parameter updates to address drifting data contexts inherent to dynamic systems. 

While several methods and approaches exist to address the multi-modal aspects of anomaly detection in geospatial data streams, the consideration of edge-feasible deployments is vastly overlooked \cite{wu2024geospatial}. Additionally, most existing works focus on anomaly identification without filtering out noise-induced flickering anomalies. In a large-scale deployment, this can result in numerous false alarms and generate large volumes of alerts. To this end, this work addresses these limitations by presenting a lightweight, computationally and communication-efficient hierarchical anomaly detection and localization framework that focuses on identifying macro anomalies in big data geospatial applications.
The contributions of this work are summarized as follows:
\begin{itemize}
\item A light-weight resource-efficient edge framework for anomaly detection in geospatial data streams.
\item The development of drill-down logic to eliminate unnecessary computations.
\item The use of a hierarchical discrete global grid partitioning system to localize the source of an anomaly at higher resolutions.
\end{itemize}

\section{System Model}
\begin{figure*}[!hbt]
\centerline{\includegraphics[width=1.35\columnwidth]{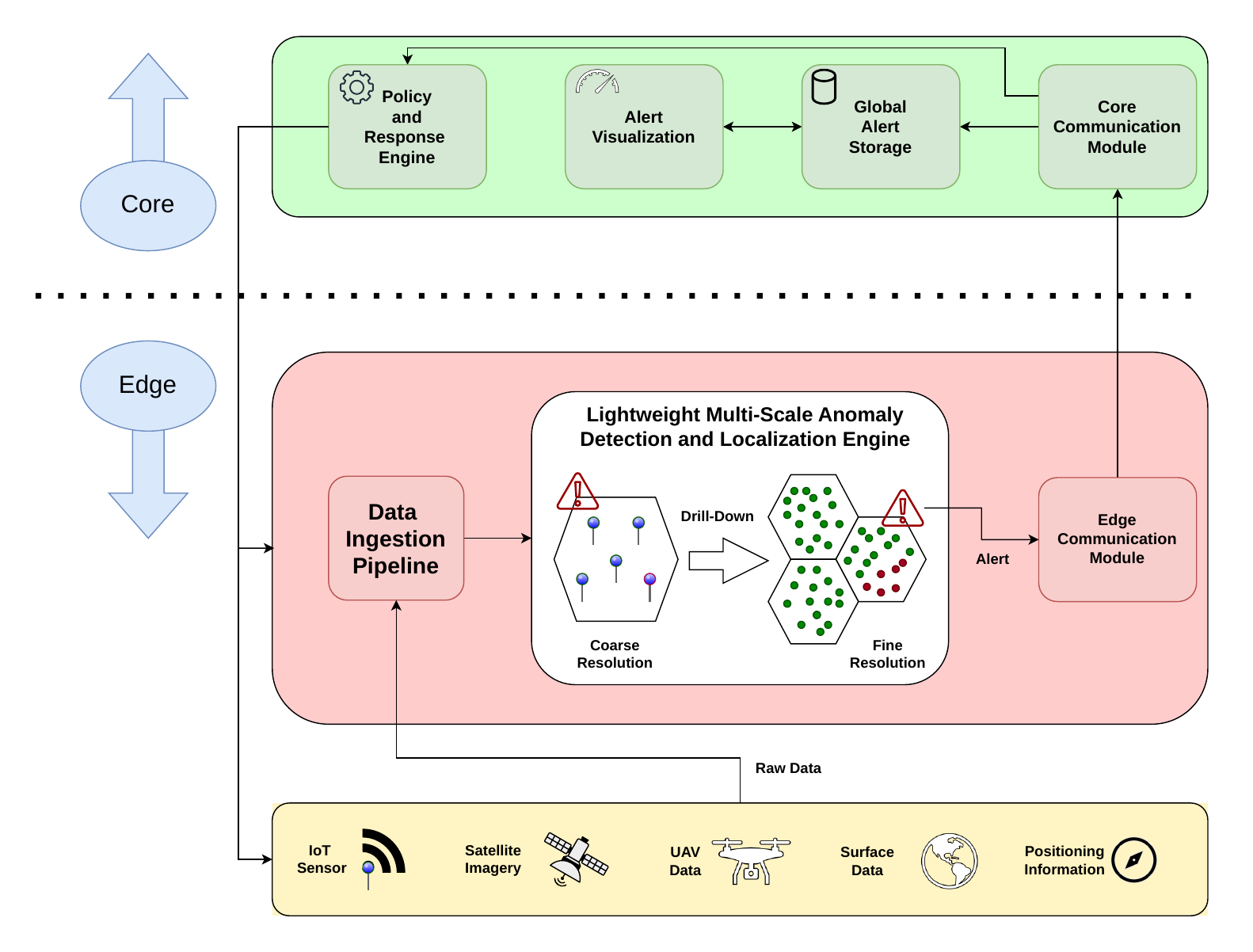}}
\caption{System Model}
\label{fig:sys_model}
\end{figure*}

A high-level overview of the system model is presented in Fig. \ref{fig:sys_model}. This model is logically separated into two regions, the edge and the core. The edge contains all sensing, processing, and analysis, whereas the core contains historical storage, visualization, and policy considerations. This logical isolation is intentional, as the core represents a centralized system (or agent), responsible for collecting information from various edge devices. It is assumed that the transfer of raw data to a centralized location is undesirable due to data privacy and resource efficiency constraints. To this end, the system effectively becomes federated, where computation is performed entirely at the edge, and insights are passed to the core for system-wide monitoring and decision-making. 

This presented system architecture is motivated by two main factors: (a) the Big Data nature of the application, and (b) the resource efficiency requirements of the domain. The remainder of this section will elaborate on these two factors and use specific characteristics of the application and domain to motivate the proposed approach.

\subsection{Big Data Motivation}
Big data applications are characterized by various properties, notably Volume, Velocity, and Variety (the 3 Vs). Beyond these three key properties, additional properties, including Veracity, Variability, and Value, also have implications on key system architectural choices and design considerations. In terms of volume, the spatial resolution of the data is a key factor. For example, the Normalized Difference Vegetation Index (NDVI), which is used as an illustrative example in this work, is collected at a spatial resolution of 1km, corresponding to approximately 813 million pixels, each encoded with attribute values and metadata. Geospatial data can be collected at various temporal resolutions, ranging from yearly to hourly, contributing to the velocity attribute. As previously mentioned, geospatial data is multi-modal and therefore exhibits high variety. Considering that sensors are a primary source of collection for this type of data, the veracity of the data in the context of sensor readings, calibration, and noise is a critical concern. Furthermore, given the high dynamicity of the system being observed, geospatial data contexts can change over time and are subject to environmental or seasonal drift events, demonstrating high variability. Finally, geospatial data is a wealth of information containing significant value with implications on critical sectors, including agriculture, meteorology, and public safety. All attributes considered, a geospatial data and intelligence pipeline meets the criteria for being classified as a big data application.

\subsection{Resource Efficiency Motivation}

From a resource efficiency perspective, significant challenges necessitate the transition to a federated and even-driven communication model, with processing at the edge. Due to the previously explained limitations regarding the storage and communication of geospatial data collection points, the transfer of large volumes of raw data must be avoided. A logical alternative is to perform data processing, intelligence, and insight generation at the edge and communication on a need-to-know basis. For example, instead of sending an entire 1km resolution with 813 million distinct pixels, preprocessing could be used to group pixels, extract aggregate statistics, identify anomalous regions, and convey areas of concern to a centralized agent. This not only saves precious communication resources, as only the relevant and actionable information is conveyed, but also directly provides actionable insights that can be leveraged as soon as they are received. Furthermore, since the amount of data being transferred is greatly reduced and only relevant insights are conveyed, the susceptibility of the data to transmission error is greatly reduced. Finally, a critical component of mission-critical and real-time applications is the Age of Information (AoI). This metric defines the freshness of information and relates to the delay between the start of data collection and the decision-making resulting from the collected data. If data is collected and processed at the edge and only small alerts need to be conveyed, small windows of communication resource availability can be leveraged to improve the AoI metric and enable real-time decision making. Additionally, edge processing reduces the processing-induced staleness of the data, further improving the AoI metric.

\section{Methodology}

Given the defined system model, this section outlines the methodological considerations of the proposed work.

\subsection{Algorithm Description}
At a high-level, the proposed algorithm takes in geospatial data, uses a discrete global grid system to aggregate individual observations, compares aggregate statistics to historical baselines and determines if anomalies are present. At its core, this work employs drill-down logic to reduce the amount of computation required to detect macro anomalies and localize their source. The drill-down logic looks for anomalous signals at low resolutions, which cover large physical locations. This approach effectively acts as a spatio-temporal low-pass filter, with high-frequency noise (flickering anomalies) being averaged out in the lower resolutions so that only significant macro anomalies are considered as the drill-down logic progresses. 

This work uses the H3 discrete global grid system \cite{Brodsky2018H3} to partition the world into hexagonal-indexed regions. The H3 partitioning has mathematical significance in the context of aggregate statistical measures and logical significance in the context of anomaly localization. Regarding the mathematical significance, each hexagonal index is surrounded by six hexagonal neighbors. The center of each hexagon is equidistant from that of its neighbor, thus reducing distortion and improving the performance of statistical and proximity-based methods. In the context of the drill-down logic itself and the localization of anomalies, each low-resolution (parent)  hexagon is composed of approximately seven higher-resolution (child) hexagons. This creates a computational flow, where if an anomaly is detected in a resolution $x$ hexagon, the resolution $x+1$ child hexagons are then evaluated to identify the source of the anomaly. This process is then repeated until a stopping criterion is met (\textit{e.g.,} target final resolution met) or the anomaly does not persist beyond a given resolution due to sampling bias.

The pseudocode for the proposed algorithm is presented in Algorithms \ref{alg:main}, \ref{alg:eval}, and \ref{alg:detect}. It should be noted that for the purposes of this work, a dataset with a yearly temporal resolution is utilized along with the H3 partitioning, for illustrative purposes; however, this work can be adapted to any level of temporal resolution and any discrete global grid system. 

Algorithm \ref{alg:main} receives all H3 hexagons belonging to the starting resolution, the year of detection, along with the threshold parameter bound as an input, and returns the identified H3 hexagonal indices that have been flagged as being anomalous. The drill-down logic is applied from the starting resolution to the max resolution. Each hexagon in the resolution in question is evaluated and an anomaly flag is determined. If an anomaly is detected, the child hexagons at the next resolution are determined and added to the list of hexagonal indices to examine.

\begin{algorithm}[h!]
\DontPrintSemicolon
\caption{Hierarchical Anomaly Detection}
\label{alg:main}
 
\KwIn{$\mathcal{G}$: cells at \Var{start\_res}, $end\_res$: max resolution, $y$: \Var{year}, $\alpha$: threshold parameter}
\KwOut{$\mathcal{F}$: cells with anomaly flags}
 
\BlankLine
$\mathcal{L} \leftarrow \mathcal{G}$;\quad $\mathcal{F} \leftarrow \emptyset$\;
\For{res $\leftarrow$ \Var{start\_res} \KwTo \Var{end\_res}}{
    \lIf{$\mathcal{L} = \emptyset$}{\textbf{break}}
    \ForEach{$g \in \mathcal{L}$}{
        $\mathcal{F} \leftarrow \mathcal{F} \cup \{$\FEval{$g$, res, \Var{y}, $\alpha$}$\}$\;
    }
    $\mathcal{A} \leftarrow \{g \in \mathcal{F} \mid \text{flag}(g) = 1,\; \text{res}(g) = \text{res}\}$\;
    \lIf{$\mathcal{A} = \emptyset$ \textbf{ or } res $= end\_res$}{\textbf{break}}
    $\mathcal{L} \leftarrow$ \FChildren{$\mathcal{A}$, res $+ 1$}\;
}
\Return $\mathcal{F}$\;
\end{algorithm}

 Algorithm \ref{alg:eval} outlines how a specific hexagonal index is evaluated. The boundary of the evaluated cell is determined using its identifier. All points falling within the geographical location of the hexagonal boundary are identified and retrieved. These points are then used to determine aggregate statistics for the hexagonal boundary, which are then compared with past historical values using statistical thresholding.

\begin{algorithm}[h!]
\DontPrintSemicolon
\caption{EvaluateCell}
\label{alg:eval}
 
\KwIn{$g$: cell, $res$: resolution, $y$: \Var{year}, $\alpha$: threshold parameter}
\KwOut{$(g,\; f: \text{flag})$}
 
\BlankLine
Retrieve points belonging to $g$ at resolution res\;
\lIf{no points or insufficient data}{\Return $(g,\;0)$}
Query annual statistics $S$ for those points\;
$f \leftarrow$ \FDetect{$S$, \Var{year}, $\alpha$}\;
\Return $(g,\; f)$\;
\end{algorithm}

Algorithm \ref{alg:detect} outlines the actual anomaly detection process. Given an H3 id and year of interest, the aggregate statistics from all previous years, starting with the first recorded year of observation to the most recent year preceding the year of interest are pulled. The mean, median, standard deviation, minimum and maximum for these statistics are then computed. Using the defined threshold parameter, $\alpha$, a mean and median threshold is constructed (\textit{e.g.,} $Thresh_{mean} = \mu \pm \alpha \sigma$). An anomaly is identified if the current year observation falls outside the mean or median threshold, or if it exceeds the established minimum or maximum values. This anomaly is then passed to Algorithm \ref{alg:main} for collection and reporting.
 
\begin{algorithm}[h!]
\DontPrintSemicolon
\caption{DetectAnomaly}
\label{alg:detect}
 
\KwIn{$S$: annual statistics, $y$: \Var{year}, threshold: $\alpha$}
\KwOut{flag $\in \{0, 1\}$}
 
\BlankLine
$\mathcal{T} \leftarrow \{S[y] \mid y \in [y_{init},\; y-1]\}$\;
\lIf{$|\mathcal{T}|$ insufficient}{\Return $0$}
Compute $\mu_v$, $\sigma_v$ for $v \in \{\bar{x},\, \tilde{x}\}$ over $\mathcal{T}$\;
Compute $[x^*_{\min},\; x^*_{\max}]$ over $\mathcal{T}$\;
\lIf{any $v$ exceeds $\mu_v \pm \alpha\,\sigma_v$, or $S[\Var{y}]$ outside $[x^*_{\min},\; x^*_{\max}]$}{\Return $1$}
\Return $0$\;
\end{algorithm}

\subsection{Complexity Analysis}

In order to understand the complexity of the proposed algorithm, it is important to note that a starting resolution of 0 contains 122 hexagons. Based on the defined algorithm, in the best-case scenario where none of the resolution 0 hexagons are anomalous, 122 hexagons will be evaluated. Each H3 hexagon is composed of approximately 7 hexagons at the subsequent resolution. Taking this into consideration, we can approximate the total number of possible evaluations as $122 \times \sum_{r=R_{min}}^{R_{max}}7^r$, representing the worst-case scenario. In the unlikely event that every hexagon at every resolution is anomalous, the drill-down logic will result in more evaluations than a flat evaluation at the highest resolution, since it has to evaluate all previous resolutions. While this worst-case scenario is more expensive than a flat evaluation of all the hexagons in the highest resolution, the probability of such a scenario materializing is very small because geospatial intelligence anomalies are spatially sparse and non-anomalous identified hexagons will not require further computation. Additionally, even if the worst-case were to materialize, the efficiencies gained through communication resource utilization due to the elimination of raw data transmission would benefit the system as a whole.

\subsection{Experiment Description}
The experiment conducted in this work to evaluate the proposed method is based on the NDVI 1 km spatial resolution yearly dataset. Data collection from this source began in 2002 and has been ongoing to the present. As an initial proof of concept, a target year of 2014 was selected, meaning that the anomaly detection algorithm will use aggregate statistics from the years [2002, 2013] to build thresholds. The selection of the year 2014 is two-fold; firstly, as the system is highly dynamic and changes over time, using excessive amounts of past data without any form of weighting to build historical behavior thresholds is susceptible to non-stationary system drift. By selecting approximately a decade, the chance of being affected by a drift event is significantly reduced. Additionally, the purpose of this work is to identify statistical macro-anomalies; however, without incorporating domain knowledge to contextualize the detected anomalies, a ground truth cannot be established. While out of the scope of this work, there exists a wealth of verifiable geophysical events, such as the California drought, which occurred in 2014 to enable a cross-reference with the raised statistical anomalies to establish a contextualized baseline performance in future work. Regarding the drill-down logic, a starting resolution of 0 was selected and a maximum resolution of 5 was used as an initial demonstration of the algorithm’s performance, with each resolution 5 hexagons corresponding to an average surface area of 253 $km^2$, representing an approximately 250:1 spatial aggregation of the NDVI observations. A threshold parameter of 3 was selected to build a threshold, representing a conservative baseline for macro-anomaly detection performance with further refinements explored in future work.

\section{Results and Analysis}
This section presents and analyzes the results. In this work the performance of the proposed algorithm is compared against a flat resolution-specific approach whereby each hexagon in each resolution is evaluated. The purpose of this comparison is to demonstrate the computational efficiency of the proposed approach, as well as the ability to filter out flickering anomalies and focus on the spatially rooted macro anomalies that persist, even when aggregating individual points over large geographical areas.

\begin{figure}[!hbt]
\centerline{\includegraphics[width=\columnwidth]{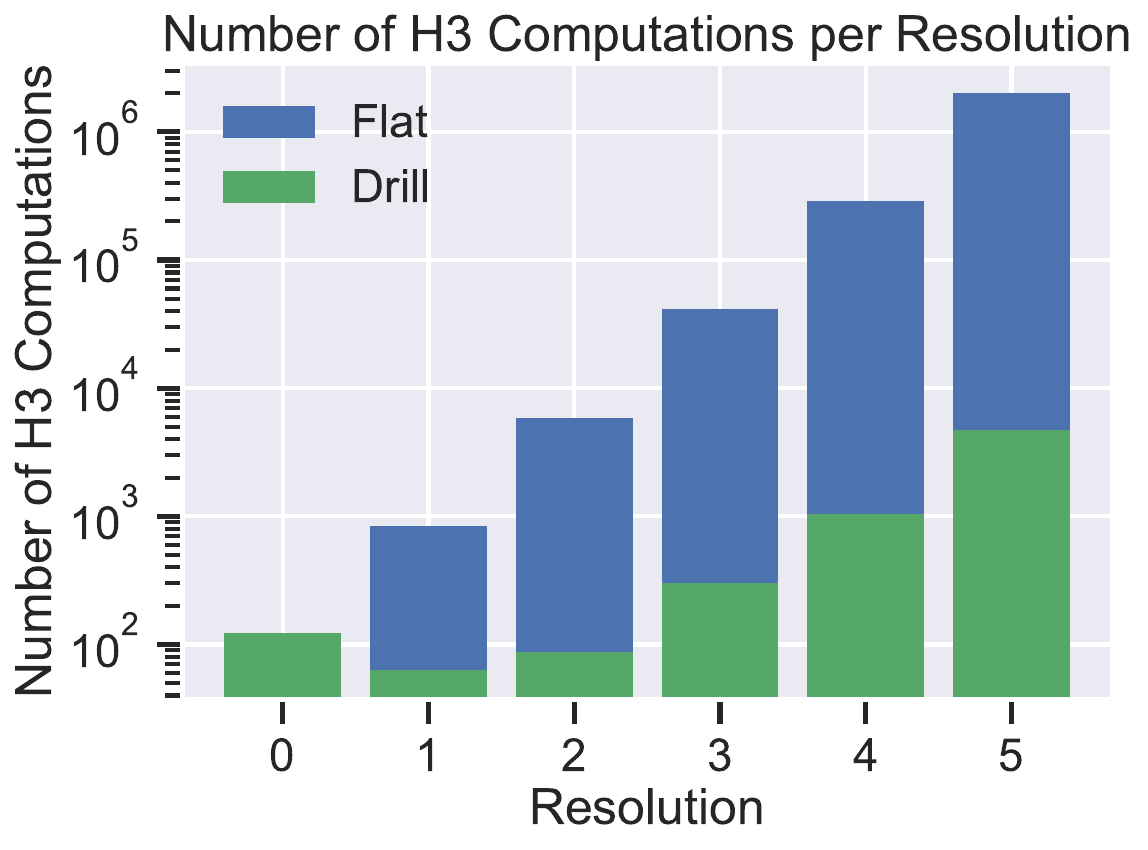}}
\caption{Computations per Resolution}
\label{fig:comps}
\end{figure}

Figure \ref{fig:comps} presents the number of computations per resolution. The flat resolution-specific approach serves as the upper bound for the number of possible hexagon evaluations. Since each H3 hexagon at a given resolution is composed of approximately seven H3 hexagons in the subsequent resolution, an exponential increase is observed in the flat approach (appearing as a linear trend on the logarithmic scale). It should be noted that a logarithmic scale has been used for visualization purposes along the y-axis. When looking at the drill-down approach, a significant reduction in the number of evaluations is observed. Consistent with the algorithm’s definition, all 122 resolution 0 hexagons are evaluated, and only those exhibiting anomalies are further expanded on in the subsequent higher resolutions. This result is consistent with the expectations of the algorithm and supports the computational filtering aspects of the approach, where a large number of low-resolution hexagons and their children do not require any further drill-down processing. The hierarchical drill-down approach achieves a computational reduction of over 99.7\% at the terminal resolution compared to the flat baseline, with approximately 4,800 hexagon evaluations as opposed to 2 million, evaluated in the flat approach. When considering the AoI metric in the broader system, the reduction in the number of evaluations directly reduces the time between data collection and anomaly reporting.

\begin{figure}[!hbt]
\centerline{\includegraphics[width=.95\columnwidth]{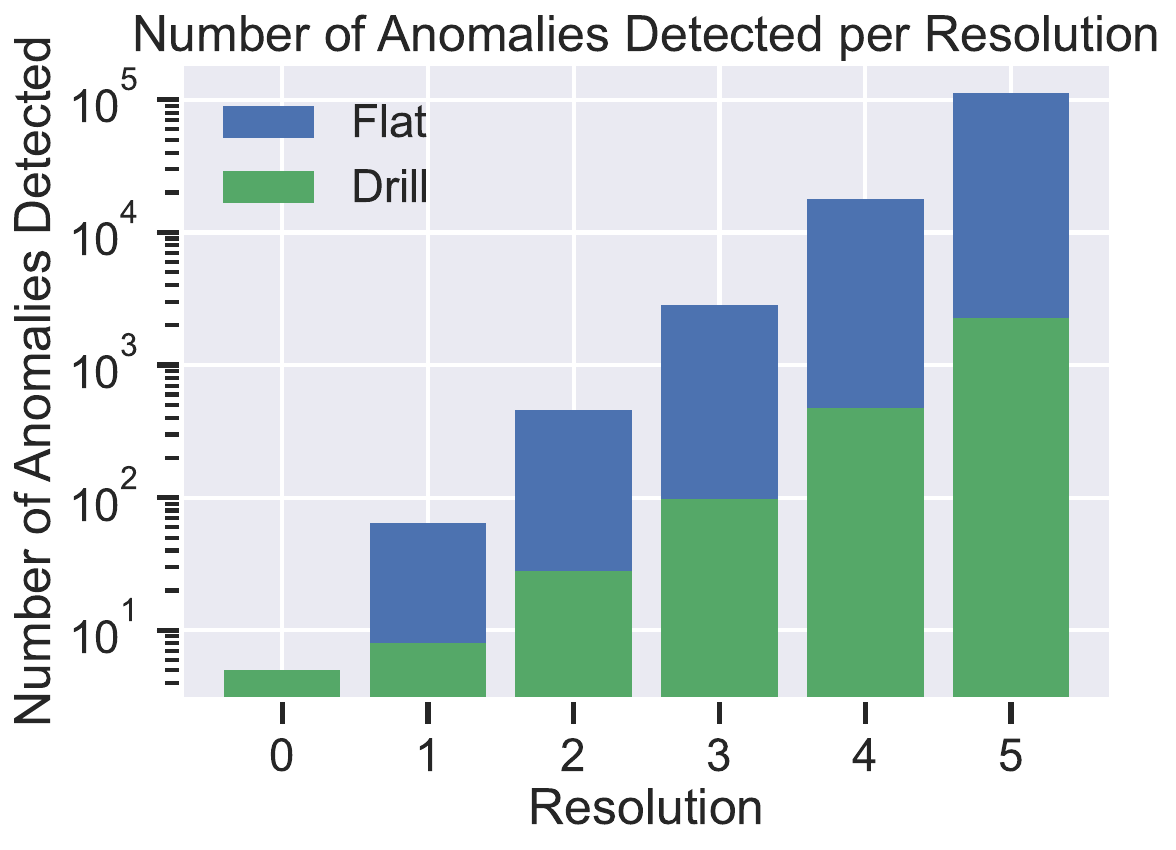}}
\caption{Anomalies per Resolution}
\label{fig:anoms}
\end{figure}

The second set of results is presented in Fig. \ref{fig:anoms} and depicts the number of anomalies flagged per resolution level. Once again, the flat resolution-specific approach serves as the upper bound for the number of anomalies detected per resolution. Since these anomalies are detected for each resolution without applying drill-down logic, there are various flickering anomalies contributing to the overall number of detections. In terms of the drill-down approach, a significant reduction in the number of anomalies is observed. This reduction has a two-fold impact; First, it reinforces the reduction in the number of evaluations per resolution, as only a small subset of the total search space requires further expansion. Second, it demonstrates that the proposed algorithm effectively prioritizes the detection of macro anomalies, by ignoring localized noise. By requiring cross-scale statistical persistence, the drill-down logic ensures that only the spatially significant anomalies are reported to the centralized agent at the core. These results demonstrate a clear enhancement to computational and communication efficiency as the initial 813 million pixels have been distilled down to a small subset of high-confidence regions.

\section{Conclusion}
The work presented in this paper proposes a lightweight multi-scale hierarchical anomaly detection and localization algorithm for big geospatial data applications at the edge. By leveraging the H3 discrete global grid system, a hierarchical partitioning of the world is used to identify anomalous regions at low resolutions and expand upon them using drill-down logic, precisely localizing regional anomalies. Using the NDVI 1km spatial resolution data as an illustrative use case, it is demonstrated that the proposed approach is computationally efficient and can effectively filter out noise and flickering anomalies, while focusing on the spatially-rooted macro anomalies present in the system. 

Future work in this area will begin by applying the proposed approach to other geospatial data sources to assess performance. These data sources will be chosen to capture various spatial and temporal resolutions to assess how the drill-down approach performs. Additionally, adaptive thresholding will be used during the drill-down approach to perform a sensitivity analysis and enable dynamic configuration of the level of filtering at the lower resolutions. Finally, a multi-layer anomaly detection model, incorporating neighborhood characteristics and intelligence methods, will be used to develop a more robust detection framework.

\section*{Acknowledgments}
This research was supported by the Defense Advanced Research Projects Agency (DARPA) [Biological Technologies Office] under Agreement No. HR0011-26-3-E036. Approved for public release; distribution is unlimited.

\bibliographystyle{IEEEtran}
\bibliography{sample}

\end{document}